\documentstyle[preprint,aps]{revtex}
\begin{document}
\draft
\title{ The effect of
quark exchange in $A=3$ mirror nuclei and neutron/proton structure
functions ratio }
\author{M. Modarres$^a$\thanks {Corresponding author, Email : modarres@khayam.ut.ac.ir},
M. M. Yazdanpanah$^b$ and F. Zolfagharpour$^a$ }
\address{$^a$Physics Department, University of  Tehran, 1439955961, Tehran Iran. \\
 $^b$Physics Department, Shahid-Ba-Honar University, Kerman, Iran.\\}
\maketitle
\begin{abstract}
 By using the quark-exchange  formalism, realistic
 Faddeev wave functions and Fermi motion effect, we investigate
 deep inelastic electron  scattering from $A=3$ mirror nuclei in the deep-valence region.
 The initial valence quarks in put are taken
 from the GRV's next-to-leading order calculations on $F_2^p(x,Q^2)$ which give very good fit to the available data
 in the $(x,Q^2)$-plane. It is shown
 that the free neutron to proton structure functions ratio can be
 extracted from corresponding EMC ratios for $^3He$ and $^3H$ mirror
 nuclei  by using self-consistent iteration procedure and the results are  in good agreement with other theoretical models
 as well as present available  experimental
 data, especially the one expected form the proposed 11 GeV
 Jefferson Laboratory.
\end{abstract}
\section{Introduction}
In the framework of {\it Standard Model}, the hadrons are composed
of valence-quarks, sea-quarks and gluons [1]. In recent years, in
order to test the perturbative and non-perturbative nature of
quantum chromodynamics, most of the experiments and theoretical
works have been focused on studies of the hadrons (mainly proton
and neutron) structure functions at small x ( x is Bjorken
scaling variable) regions [2] where the sea-quarks and gluons play
important roles. While the middle x region i.e. $0.3<x<0.7$ at
moderate $Q^2$ (the photon 4-momentum), have been assumed to be
understood and can be explained by the valence-quarks dynamics,
there is lake of information about the nucleons structure
functions in the deep-valence region, $x>0.7$  at any $Q^2$.
There are a few reasons to study the quark distributions in
nucleons at  large x. The d/u quark distribution function ratio
near $x\simeq 1$ can give information about (1) the spin-flavor
symmetry breaking in the nucleons (2) the onset of perturbation
behavior [3] and (3) search for new physics beyond {\it Standard
Model} [4] in the high energy colliders at high $Q^2$ e.g. the
uncertainty in the  gluon distributions.

While the proton structure functions is quite well known both
experimentally  and theoretically [1-3], the neutron structure
functions usually  should be extracted from deuterium and because
of the large nuclear corrections, there could be uncertainties as
large as $\%50$ in the d/u or $F^n_2/F^p_2\simeq{1+4{d\over
u}\over 4+{d\over u}}$ ratios [5]. The bounds value of ${1\over 4
}\leq { F^n_2\over F^p_2}\leq 4 $ has been  proposed by Nachtmann
[6] for the all x regions. On the other hand, based on different
models the values of ${2\over 3}$ (SU(6) symmetry)[7], ${1\over
4}$ (phenomenological and Regge considerations)[8 ] and ${3\over
7}$ (quarks counting roles and perturbative QCD) [9] have been
predicted  as $x\longrightarrow 1$.

Recently, the possible use of an unpolarized tritium target has
been proposed, by using the $11 GeV$ upgraded beam of Jefferson
Laboratory [10], and aimed to measure $F^n_2$, using the ratio of
structure functions of helium-3 $({\cal F}^{^3He}_2)$ and tritium
$({\cal F}^{^3H}_2)$ in order to reduce the systematic errors both
in the experimental measurements and theoretical calculations
(which are model dependent).

Several years ago the quark-exchange formalism was originally
introduced by Hoodbhoy and Jaffe [HJ] to investigate the quarks
distributions in nuclear systems [11,12]. This formalism was
applied by one of the authors (MM) to light nuclei [13] and
nuclear matter [14] and was reformulated by us to derive the spin
structure function of the three-nucleon systems as well as the
proton and neutron [15] and finally it was used as the initial
conditions for the QCD evolution equations [16] to calculate the
sea-quarks and gluons contributions to the proton structure
function at the leading and next-to-leading order (NLO) levels.
In the most of the above calculations we found satisfactory
agreements with the available experimental data. So we think the
quark-exchange formalism is good motivation for looking at
$F^n_2/F^p_2$ ratios at deep-valence region i.e. $x\geq 0.7$.

Recently, several groups have paid attention to the above matter
[17-22] and have used different models (mostly based on impulse
formalism  and different spectral function  approximation) to
calculate $F^n_2/F^p_2$ ratios at deep-valence region. We will
present their results in this work and compare them with ours.

So the paper will be organized as follows: In section II we
briefly explain the quark-exchange model and we calculate the
valence quark momentum distributions for proton and neutron in
$^3He$ and $^3H$. In section III , we calculate the structure
functions of helium 3 and tritium by including the Fermi motion
effect. The self-consistent calculation (iteration procedure)  of
the ratio $F^n_2/F^p_2$ will be explained in section IV. Finally
in section V we present our numerical results, discussion and
conclusion.
\section{The Quark-Exchange Formalism}
Let us start with a brief summary of the quark-exchange formalism.
We take the nucleon states to be composed of three valence quarks
[11,15],
\begin{equation}
| {\alpha} {\rangle}={\cal N}^{{\alpha}^{\dagger}}| 0 {\rangle}
={1 \over \sqrt{3!}}{\cal N}^{\alpha}_{{\mu}_1 {\mu}_2 {\mu}_3}
q^{\dagger}_{{\mu}_1}q^{\dagger}_{{\mu}_2}
q^{\dagger}_{{\mu}_3} | 0 {\rangle}
\end{equation}
where $ \alpha$ ($\mu_i$) describe the nucleon (quark) states
${\{\vec P}, M_S ,M_T \}$ ( ${\{\vec k} , m_s , m_t , c\}$)(note
that $M_T(m_t)=+{1\over 2 }$ and $-{1\over 2}$ for the proton
(up-quark) and neutron (down-quark), respectively). As usual
$q^\dagger_\mu$ ($\cal N^{{\alpha}^{\dagger}}$) denote the
creation operators for the quarks (nucleons) with the state index
$\mu$ ($\alpha$). With this convention that a repeated index
means a summation as well as integration over $\vec k$. The
totally antisymmetric nucleon wave function ${\cal
N}^{\alpha}_{{\mu}_1 {\mu}_2 {\mu}_3}$ are written as,
\begin{equation}
{\cal N}^{\alpha}_{{\mu}_1 {\mu}_2 {\mu}_3}=
D(\mu_1,\mu_2,\mu_3;\alpha_i)
\times
{ \delta({\vec k_1}+{\vec k_2}+{\vec k_3}-{\vec P}})
{\phi ({\vec k_1},{\vec k_2},{\vec k_3},{\vec P})}
\end{equation}
where ${\phi ({\vec k_1},{\vec k_2},{\vec k_3},{\vec P})}$ is the
three nucleon wave function which  is approximated by  a Gaussian form
(b $\simeq$ nucleons radius) :
\begin{equation}
\phi ({\vec k_1},{\vec k_2},{\vec k_3},{\vec P})=
{\left(\frac {3b^4}{{\pi}^2}\right)^{3 \over 4}}
\exp[-b^2(\frac {(k_1^2+k_2^2+k_3^2)}{2}+\frac {{b^2}{P^2}}{6}]
\end{equation}
$D(\mu_1,\mu_2,\mu_3;\alpha_i)$ are the product of four Clebsch
-Gordon coefficients, $C^{{j_1} {j_2} {j}}_{{m_1} {m_2} {m}}$,
($\epsilon_{{c_1}{c_2}{c_3}}$ are the color factors) which are
defined as,
\begin{equation}
D(\mu_1,\mu_2,\mu_3;\alpha_i) = {1 \over \sqrt{3!}}
\epsilon_{c_1c_2c_3}
{1 \over \sqrt 2}\sum_{s,t=0,1}
C^{{1\over 2}  s  {1 \over 2}}_{m_{s_\sigma} m_s
M_{S_{\alpha_i}}}
C^{{1 \over 2} { 1 \over 2}  s}_{m_{s_\mu}
m_{s_\nu} m_s}C^{{1 \over 2}  t
{1 \over 2}}_{m_{t_\sigma} m_t M_{T_{\alpha_i}}}
C^{{1 \over 2}{ 1 \over 2} t}_{m_{t_\mu} m_{t_\nu} m_t}
\end{equation}

Now, based on the nucleon creation operators, we can define the
nucleus states as,
\begin{equation}
|{\cal A}_i=3 {\rangle}={(3!)^{-{1 \over 2}}}
{\chi}^{{\alpha}_1{\alpha}_2{\alpha}_3}{\cal N}^{{{\alpha}_1}^{\dagger}}
{\cal N}^{{{\alpha}_2}^{\dagger}}{\cal N}^{{{\alpha}_3}^{\dagger}}|0{\rangle}
\end{equation}
where ${\chi}^{{\alpha}_1{\alpha}_2{\alpha}_3}$ are a complete
antisymmetric nuclear wave functions (they are  taken from the
Faddeev calculation with Reid soft core potential [15]. According
to Afnan et al. [20], the choice of nucleon-nucleon potential
does not affect the EMC result. However we will investigate this
matter in our future works) and it should be interpreted as the
center of mass motion of the three nucleons. By using the same
definition as the one we did for C-G coefficients in equation (4)
i.e.,
\begin{equation}
D(\alpha_1,\alpha_2,\alpha_3;{\cal A}_i)=
{1 \over \sqrt 2}{\sum_{S,T=0,1}}
C^{{1\over 2} S {1 \over 2}}_{M_{S_{\alpha_1}} M_S M_{S_i}}
C^{{1 \over 2} {1 \over 2}  S}_{M_{S_{\alpha_2}} M_{S_{\alpha_3}}
{\cal M_S}}
C^{{1 \over 2} T {1 \over 2}}_{M_{T_{\alpha_1}} M_T M_{T_i}}
C^{{1 \over 2} {1 \over 2} T}_{M_{T_{\alpha_2}} M_{T_{\alpha_3}}
{\cal M_T}}
\end{equation}
Then we can write the nuclear wave functions as
\begin{equation}
{\chi}^{{\alpha}_1{\alpha}_2{\alpha}_3}={\chi(\vec P,\vec q)}
D({\alpha}_1,{\alpha}_2,{\alpha}_3;{\cal A}_i)
\end{equation}

Relevant information comes from the momentum distribution of the
constituent quarks, which can be defined for valence quarks with
the fixed flavor and nucleon iso-spin projection in the three
nucleon system as,
\begin{equation}
\rho_{m_t}^{M_T}(\vec k;{\cal A}_i)={{\langle} {\cal A}_i=3
|q^{\dagger}_{\bar {\mu}}q_{\bar {\mu}} |{\cal A}_i=3{\rangle}
\over {\langle}{\cal A}_i=3 |{\cal A}_i=3{\rangle}}
\end{equation}
The sign bar means no summation on $M_T$, $m_t$ and  integration
over $\vec k$ on the repeated index $\mu$. The calculation of
${\langle}{\cal A}_i=3 |{\cal A}_i=3{\rangle}$ would become
straightforward by doing summation over $\bar \mu$,
$${\langle}{\cal A}_i=3 |{\cal A}_i=3{\rangle}={1 \over 9}
{{\langle}{\cal A}_i=3|q^{\dagger}_{ {\mu}}q_{{\mu}} |{\cal A}_i}=3
{\rangle}$$
$$=\chi^{* \alpha_1 \alpha_2 \alpha_3 } ( \delta^{\alpha_1 \beta_1}
\delta^{\alpha_2 \beta_2} \delta^{\alpha_3 \beta_3}
-{\cal E}^{\alpha_1 \alpha_2 \alpha_3 ,\beta_1 \beta_2 \beta_3}_{\mu \mu})
{\chi}^{\beta_1 \beta_2 \beta_3}$$
where
\begin{equation}
{\cal E}^{{\alpha}_1{\alpha}_2{\alpha}_3,{\beta}_1{\beta}_2{\beta}_3}_{{\mu}
{\mu}}={\cal N}^{{\alpha}_2}_{{\mu}_1{\mu}_2{\mu}_3}
{\cal N}^{{\beta}_2}_{{\mu}_2{\mu}_3{\rho}_1}
{\cal N}^{{\alpha}_3}_{{\rho}_1{\rho}_2{\rho}_3}
{\cal N}^{{\beta}_3}_{{\mu}_1{\rho}_2{\rho}_3}
{\delta}^{{\alpha}_1{\beta}_1}
\end{equation}

After performing some algebra, one could drive the following equation
for the expectation value of
$q^\dagger q$,
$${\langle}{\cal A}_i=3|q^\dagger_{\bar \mu} q_{\bar \mu}
|{\cal A}_i=3{\rangle}=9\chi^{* \alpha_1 \alpha_2 \alpha_3}
({\cal U}^{\alpha_1 \alpha_2 \alpha_3 ,
 \beta_1 \beta_2 \beta_3}_{\bar \mu \bar \mu}
-{\cal V}^{\alpha_1 \alpha_2 \alpha_3,
\beta_1 \beta_2 \beta_3}_{\bar \mu \bar \mu})
\chi^{\beta_1 \beta_2 \beta_3}$$
where
\begin{equation}
{\cal U}^{{\alpha}_1{\alpha}_2{\alpha}_3,
{\beta}_1{\beta}_2{\beta}_3}_{{\bar \mu}
{{\bar \mu}}}={\cal N}^{{\alpha}_1}_{{\bar \mu}{\sigma}_2{\sigma}_3}
{\cal N}^{{\beta}_1}_{{\bar {\mu}}{\sigma}_2{\sigma}_3}
{\delta}^{{\alpha}_2{\beta}_2}
{\delta}^{{\alpha}_3{\beta}_3}
\end{equation}
and

$${\cal V}^{\alpha_1\alpha_2\alpha_3,
\beta_1\beta_2\beta_3}_{\bar \mu
\bar \mu}=3{\cal N}^{\alpha_1}_{\bar \mu\sigma_2\sigma_3}
{\cal N}^{\beta_1}_{\bar \mu\sigma_2\sigma_3}
{\cal N}^{\alpha_2}_{\mu_1\mu_2\mu_3}
{\cal N}^{\beta_2}_{\rho_1\mu_2\mu_3}
{\cal N}^{\alpha_3}_{\rho_1\rho_2\rho_3}
{\cal N}^{\beta_3}_{\mu_1\rho_2\rho_3}$$
\begin{equation}
+4{\cal N}^{\alpha_2}_{\bar \mu\mu_1\mu_2}
{\cal N}^{\beta_2}_{\bar \mu\mu_2\rho_1}
{\cal N}^{\alpha_3}_{\rho_1\rho_2\rho_3}
{\cal N}^{\beta_3}_{\mu_1\rho_2\rho_3}
\delta^{\alpha_1\beta_1}
+2{\cal N}^{\alpha_2}_{\mu_1\mu_2\mu_3}
{\cal N}^{\beta_2}_{\bar \mu\mu_2\mu_3}
{\cal N}^{\alpha_3}_{\bar \mu\rho_2\rho_3}
{\cal N}^{\beta_2}_{\mu_1\rho_2\rho_3}\delta^{\alpha_1\beta_1}
\end{equation}

Then by assuming the nucleus to be in the rest frame and defining
the Fourier transform of $\chi (\vec P,\vec q)$, we can calculate
the expectation values of different terms in equations (10) and
(11)

$$\chi^{*\alpha_1\alpha_2\alpha_3}
{\cal N}^{\alpha_1}_{\bar \mu\sigma_2\sigma_3}
{\cal N}^{\beta_1}_{\bar \mu\sigma_2\sigma_3}
\delta^{\alpha_2\beta_2}
\delta^{\alpha_3\beta_3}\chi^{\beta_1\beta_2\beta_3}
=\left({3b^2 \over 2\pi^2} \right)^{3 \over 2}
exp[-{ 3 \over 2}b^2{\vec k}^2]
D(\bar \mu,\sigma_2,\sigma_3;\alpha_1)$$
\begin{equation}
D(\bar \mu,\sigma_2,\sigma_2,\sigma_3;\beta_1)
D(\alpha _1,\alpha _2,\alpha _3;{\cal A}_i)
D(\beta _1,\beta _2,\beta _3;{\cal A}_i)
\delta^{\alpha _2\beta _2}
\delta^{\alpha _3\beta _3}
\end{equation}

$$\chi^{*\alpha_1\alpha_2\alpha_3}
{\cal N}^{\alpha_1}_{\bar \mu\sigma_2\sigma_3}
{\cal N}^{\beta_1}_{\bar \mu\sigma_2\sigma_3}
{\cal N}^{\alpha_2}_{\mu_1\mu_2\mu_3}
{\cal N}^{\beta_2}_{\rho_1\mu_2\mu_3}
{\cal N}^{\alpha_3}_{\rho_1\rho_2\rho_3}
{\cal N}^{\beta_3}_{\mu_1\rho_2\rho_3}
\chi^{\beta_1\beta_2\beta_3}=$$
$$I \left({27b^2 \over 8\pi^2}\right)^{ 3 \over 2}
exp[-{3 \over 2}b^2{\vec k}^2]
D(\bar \mu,\sigma_2,\sigma_3;\alpha_1)
D(\bar \mu,\sigma_2,\sigma_3;\beta_1)
D(\mu_1,\mu_2,\mu_3;\alpha_2)
D(\rho_1,\mu_2,\mu_3;\beta_2)$$
\begin{equation}
D(\rho_1,\rho_2,\rho_3;\alpha_3)
D(\mu_1,\rho_2,\rho_3;\beta_3)
D(\alpha _1,\alpha _2,\alpha _3;{\cal A}_i)
D(\beta _1,\beta _2,\beta _3;{\cal A}_i)
\end{equation}

$${\chi}^{*{\alpha}_1{\alpha}_2{\alpha}_3}
{\cal N}^{{\alpha}_2}_{{\bar {\mu}}{\mu}_1{\mu}_2}
{\cal N}^{{\beta}_2}_{{\bar {\mu}}{\mu}_2{\rho}_1}
{\cal N}^{{\alpha}_3}_{{\rho}_1{\rho}_2{\rho}_3}
{\cal N}^{{\beta}_3}_{{\mu}_1{\rho}_2{\rho}_3}
{\delta}^{{\alpha}_1{\beta}_1}  {\chi}^{{\beta}_1{\beta}_2{\beta}_3}=$$
$$I {\left(27b^2 \over 7\pi^2\right)}^{3 \over 2}
exp[-{12 \over 7}b^2 {\vec k}^2]
D({\mu}_1,{\mu}_2,{\bar {\mu}};{\alpha}_2)
D({\rho}_1,{\mu}_2,{\bar {\mu}};{\beta}_2)
D({\rho}_1,{\rho}_2,{\rho}_3;{\alpha}_3)$$
\begin{equation}
D({\mu}_1,{\rho}_2,{\rho}_3;{\beta}_3)
D({\alpha }_1,{\alpha }_2,{\alpha }_3;{\cal A}_i)
D({\beta }_1,{\beta }_2,{\beta }_3;{\cal A}_i)
{\delta}^{{\alpha }_1{\beta }_1}
\end{equation}

$$\chi^{*\alpha_1\alpha_2\alpha_3}
{\cal N}^{\alpha_2}_{\mu_1\mu_2\mu_3} {\cal N}^{\beta_2}_{\bar
\mu\mu_2\mu_3} {\cal N}^{\alpha_3}_{\bar \mu\rho_2\rho_3} {\cal
N}^{\beta_2}_{\mu_1\rho_2\rho_3}\delta^{\alpha_1\beta_1}
\chi^{\beta_1\beta_2\beta_3}=$$
$$I \left({27b^2 \over 4\pi^2}\right)^{3 \over 2}exp[-{3b^2 \vec k^2}]
D(\mu_1,\mu_2, \mu_3;\alpha_2)D(\mu_2,\mu_3,\bar \mu;\beta_2)
D(\bar \mu,\rho_2,\rho_3;\alpha_3)$$
\begin{equation}
D(\mu_1,\rho_2,\rho_3;\beta_3)
D(\alpha _1,\alpha _2,\alpha _3;{\cal A}_i)
D(\beta _1,\beta_2,\beta _3;{\cal A}_i)
\delta^{\alpha _1\beta _1}
\end{equation}

$$\chi^{*\alpha_1\alpha_2\alpha_3}
 {\cal N}^{\alpha_2}_{\mu_1\mu_2\mu_3}
{\cal N}^{\beta_2}_{\mu_2\mu_3\rho_1} {\cal
N}^{\alpha_3}_{\rho_1\rho_2\rho_3} {\cal
N}^{\beta_3}_{\mu_1\rho_2\rho_3}\delta^{\alpha_1\beta_1}
\chi^{\beta_1\beta_2\beta_3} =$$
$$I{({3 \over 2})}^3
D(\mu_1,\mu_2,\mu_3;\alpha_2)D(\rho_1,\mu_2,\mu_3;\beta_2)
D(\rho_1,\rho_2,\rho_3;\alpha_3)$$
\begin{equation}
D(\mu_1,\rho_2,\rho_3;\beta_3)
D(\alpha _1,\alpha _2,\alpha _3;{\cal A}_i)
D(\beta _1,\beta _2,\beta _3;{\cal A}_i)
\delta^{\alpha_1\beta _1}
\end{equation}
where
\begin{equation}
I=8 \pi^2 \int_0^{\infty} x^2 \ dx \int_0^{\infty} y^2 \ dy
\int_{-1}^{1} \ d(cos\theta)
exp[-{3x^2 \over 4b^2}]|\chi(x,y,cos\theta)|^2
\end{equation}
The above  equations have been calculated with the same
approximation as the one used in the references [11,16], specially
a leading order expansion for $\chi (\vec P,\vec q)$ [16]. AS we
pointed out  before according to references [20,23], the other
choice of nucleon-nucleon potentials do not change the ${\cal
A}=3$ nuclear wave function and the EMC effect very much .
\section{NUCLEUS STRUCTURE FUNCTION}
The structure function measures the distribution of quarks as a
function of $k^+$ (the light-cone momentum of initial quark) in
the target rest fame which is equivalent to boosting the nucleus
to an infinite momentum fame. This is usually done by using an
{\it ad hoc} prescription for $k^0$ as a function of $|{\vec
k}|$($k^0=[({\vec k}^2+m^2)^{1\over 2}-\epsilon_0]$). It has been
shown that the resulting structure functions are  not sensitive
to this assumption [11,24]. So the valence-quark distribution at
each $Q^2$, can be related to momentum distribution for each
flavor in the nucleons of nucleus  ${\cal A}_i$ according to the
following equation ($j=p,n$ ($a=u,d$) for protons (up-quarks) and
neutrons (down-quarks), respectively),
\begin{equation}
{q_j}^a(x,Q^2;{\cal A}_i)=\int \rho^j_a({\vec k};{\cal A}_i)
\delta(x-{k_+ \over M}) \ d{\vec k}
\end{equation}
After doing the angular integration, we get,
\begin{equation}
{q_j}^a(x,Q^2;{\cal A}_i)={2\pi M}  {\int_{k_{min}}^{\infty}}
\rho^j_a({\vec k};{\cal A}_i)k \ dk
\end{equation}
with
\begin{equation}
k_{min}(x)={(xM +\epsilon_0)^2-m^2 \over 2(xM+\epsilon_0)}
\end{equation}
where m (M) is the quark (nucleon) mass and $\epsilon_0$ is the
quark binding energy. For each $Q^2$ value, it is  possible (by
using the fitting procedure as it will be described later on) to
calculate the corresponding values of m and $\epsilon_0$.
Finally, the target structure function $F_2^{{\cal A}_i}(x,Q^2)$
can be expressed in terms of the valence quark distributions as
following :
\begin{equation}
F_{2,ex}^{{\cal A}_i}(x,Q^2)=x\sum_{a=u,d; j=p,n}
Q_a^2{q_j}^a(x,Q^2;{\cal A}_i)
\end{equation}
So we have found the relation between quark momentum distribution
and the target structure function. In order to fix the values of m
and $\epsilon_0$, we apply the above equations as well as the
quark-exchange formalism  to the proton $({\cal A}_i=1,{\cal
M}_T={\cal M}_T ={1\over 2})$ as our target (obviously by
considering the proton as our target there is no exchange term
and we have just the direct term). In this case for each value of
b we find the pairs m and $\epsilon_0$ such that we get the best
fit to the valence u and d quarks distribution functions of GRV's
partons structure function calculations [25]. The GRV's partons
structure functions fit the experimental proton structure
 function data over the whole range of $(x,Q^2)$ plane very well. Figure 1 shows our
 fitted proton structure function (obviously only for valence
 quarks)
for $b=0.8 fm$ with the $(m,\epsilon_0)$ pairs of
$(165MeV,150MeV)$ and $(165MeV,250MeV)$  at $Q^2=4 GeV^2$. The
dotted curve is those of GRV and other curves have been obtained
by the same pairs of $(m,\epsilon_0)$, as above, but for
different b. The experimental data of SLAC [19] and NMC [29] as
well as  the full  GRV's NLO structure functions for proton (dash
curve)  and neutron  (dash dotted curve) have been also given for
comparison. It is seen that for $x\geq 0.2$ we get very good fit
to the data. With the same parameters we can calculate the
corresponding neutron, $^3He$ and $^3H$ structure functions which
are uncertain because of the lake of information about the
neutron structure functions. Obviously in general our method
fails for $x\rightarrow 0$ (because of fitting procedure). It is
also not good as $x\rightarrow 1$
 especially  for the nucleus target, because we have ignored the Fermi motion effect by
 using the expansion in equation (17). In order to take into the
 account the Fermi motion effect we  have worked in the convolutions approach and   the harmonic
 oscillator basis for tritium and helium 3 with the procedure
 described in reference [1,26] as following. In the convolutions
 approach the nucleus structure function can be written as :
\begin{equation}
F_2^{{\cal A}_i}(x,Q^2)=\sum_{j=p,n}\int_x^\infty dz{f_j}^{{\cal
A}_i}(z)F_2^j(x/z,Q^2)
\end{equation}
where ${f_j}^{{\cal A}_i}(z)$ and $F_2^j(x,Q^2)$ are the proton or
neutron distribution  functions in the target nucleus and the
corresponding structure functions, respectively. In the harmonic
oscillator basis we have:
\begin{equation}
{f_j}^{{\cal A}_i}(z)=\sum_{n_j,l_j}{\cal G}_{n_j,l_j}{\cal
S}_{n_j,l_j}(z,M,\hbar\omega,\varepsilon_{n_jl_j})
\end{equation}
where ${\cal G}_{n,l}$, ${\cal
S}_{n,l}(z,M,\hbar\omega,\varepsilon_{nl})$,
$\hbar\omega={\hbar^2\alpha^2\over M}$ and $\varepsilon_{nl}$ are
the occupation numbers,  sum of harmonic oscillators polynomials
(for present calculation it is just a Gaussian function),
oscillator parameter and single particle energies [26]. For
three-body system $\alpha^2={9\over 2<r^2>}$ where the rms radius
,$(<r^2>)^{1\over 2}$, is $1.95fm$ and  $1.7fm$ for helium 3 and
tritium [27], respectively. We also set $\varepsilon_{nl}=0$,
since we only intend  to calculate the Fermi motion effect (the
quark exchange is responsible for the binding effect). Figure
2(a) shows the ratio of Fermi motion effect for $^3He$ to $^3H$
i.e ${\cal R} ^{^3He}_{Fermi}(x,Q^2)/{\cal
R}^{^3H}_{Fermi}(x,Q^2)$ where,
\begin{equation}
{\cal R}_{EMC}^{^3He,Fermi}(x,Q^2)={{\cal
F}^{^3He}_{Fermi,2}(x,Q^2)\over 2F_2^p(x,Q^2)+F_2^n(x,Q^2)}, \
{\cal R}_{EMC}^{^3H,Fermi}(x,Q^2)={{\cal
F}^{^3H}_{Fermi,2}(x,Q^2)\over F_2^p(x,Q^2)+2F_2^n(x,Q^2)}.
\end{equation}
It is seen from this figure  that the Fermi motion is the same for
helium 3 and tritium up to $x\simeq 0.55$ and since the size of
$^3He$ is larger than $^3H$, the ratio start to decrease from this
value i.e. the Fermi motion effect has larger size in tritium with
respect to helium,  as one expects.

Finally the total structure function for each nucleus can be
written as the sum of the quark-exchange, equation (21) and Fermi
motion, equation (22), effects :
\begin{equation}
{\cal F}^{{\cal A}_i}_2(x,Q^2)=F^{{\cal
A}_i}_{2,ex}(x,Q^2)+F^{{\cal A}_i}_{2,Fermi}(x,Q^2)
\end{equation}

We hope in our future works we  omit this approximation by
calculating the Fermi motion effect in the framework of
quark-exchange formalism and the Faddeev wave functions for
three-nucleon systems.
\section{Self-consistent treatment of $F^n_2/F^p_2$}
By defining the EMC-type ratios for the structure functions of
helium 3 and  tritium (as  the one we did for the Fermi motion
effect) i.e.:
\begin{equation}
{\cal R}_{EMC}^{^3He}(x,Q^2)={{\cal F}^{^3He}_2(x,Q^2)\over
2F_2^p(x,Q^2)+F_2^n(x,Q^2)}
\end{equation}
and
\begin{equation}
{\cal R}_{EMC}^{^3H}(x,Q^2)={{\cal F}^{^3H}_2(x,Q^2)\over
F_2^p(x,Q^2)+2F_2^n(x,Q^2)}
\end{equation}
we can calculate the above   EMC ratios by using the  fitted
GRV's proton and neutron structure functions, the convolutions
approach and the quark-exchange formalism. In figures 2(b) and
2(c) we have plotted these ratios for different values of b. The
HERMES helium 3 data is taken form reference [28] at $Q^2\simeq 7
GeV^2$. We should point out here that the quoted  data is the
combination of helium 3,  deuterium and proton cross-sections (see
Ackerstaff [28]) (${\cal R}={{\cal F}^{^3He}_2\over {\cal
F}^{d}_2+F^p_2}$). We get good agreement with the present
available data, especially for the deep-valence  region. It is
interesting that in this region the EMC ratios are not very
sensitive to the different values of b.

Now by dividing the above two EMC ratios  we find the following
equation:
\begin{equation}
{\cal R}_{EMC}^{^3He/{^3H}}(x,Q^2)={{\cal
R}_{EMC}^{^3He}(x,Q^2)\over {\cal R}_{EMC}^{^3H}(x,Q^2)}={\cal
R}^{^3He/{^3H}}(x,Q^2)[{{1+2F_2^n(x,Q^2)/F_2^p(x,Q^2)}\over
2+F_2^n(x,Q^2)/F_2^p(x,Q^2)}]
\end{equation}
with
\begin{equation}
{\cal R}^{^3He/{^3H}}(x,Q^2)={{\cal F}^{^3He}_2(x,Q^2)\over {\cal
F}^{^3H}_2(x,Q^2)}
\end{equation}
and  by imposing the constrain,
\begin{equation} {\cal
R}_{EMC}^{^3He/{^3H}}(x,Q^2)\simeq 1
\end{equation}
which will be discussed later on.

The above equation can be solved for the neutron to proton
structure functions ratio, $F_2^n(x,Q^2)/F_2^p(x,Q^2)$, in terms
of the EMC ratio, ${\cal R}_{EMC}^{^3He/{^3H}}(x,Q^2)$, which
directly yields :
\begin{equation}
{F_2^n(x,Q^2)\over F_2^p(x,Q^2)}= {2{\cal
R}_{EMC}^{^3He/{^3H}}(x,Q^2)-{\cal R}^{^3He/{^3H}}(x,Q^2)\over
2{\cal R}^{^3He/{^3H}}(x,Q^2)-{\cal R}_{EMC}^{^3He/{^3H}}(x,Q^2)}
\end{equation}
Equations (28) and (31) are coupled in terms of the neutron to
proton  structure function ratio and will be solved by  a
self-consistent iteration procedure i.e the out put of equation
(31) will be used as the  in put for equation (28) and  so on, by
using  equation (30) as our constrain.
\section{result, discussion and conclusion}
In figure 3  the ratios of structure functions of helium 3 to
tritium (equation (29)) have been plotted for different values of
b (full curves) by using GRV's structure functions.   The
sun-burst points are the expected ratio which have been estimated
by using the kinematics of the proposed 11 GeV Jefferson
Laboratory experiment [10,20]. It is seen that there is  very
good agreement between the estimated prediction [10,20] and the
present calculation. This shows that the GRV's proton and neutron
structure functions and present model without any parameter can
predict reasonabley the structure functions of helium 3 and
tritium.

The EMC ratios (equation (28)) has been given in figure 4. In
general, since we have taken into the account all of the
properties of the structure functions of proton, neutron, helium
3 and tritium, it is expected that ${\cal
R}_{EMC}^{^3He/{^3H}}(x,Q^2)\simeq 1$, as we imposed it as a
constrain in equation (30). However it is seen  that the
calculated EMC ratio has a very small variation from ${\cal
R}_{EMC}^{^3He/{^3H}}(x,Q^2)=1$. But this deviation increases as
$x\rightarrow 1$. The  similar behavior is also seen in the others
calculations in which the impulse approximation have been used
[20].

Finally, in figure 5 the calculated neutron to proton structure
functions ratios have been plotted for different values of b. The
curve without iteration is the one in which we have not imposed
the constrain equation (30). The first and second iteration
curves have been calculated by considering the constrain ${\cal
R}_{EMC}^{^3He/{^3H}}(x,Q^2)=1.01$. The third iteration is not
distinguishable from the second one. The limit values of this
ratio for $x=1$ (as it was discussed  in the introduction) from
different models are shown in the left part of this figure by
arrows. The data are from references [17,18,19,28]. The iterations
approximately converge after the third iteration. Our results
cover all of the range of   present available data. This can be
done by varying ${\cal R}_{EMC}^{^3He/{^3H}}(x,Q^2)$ between 0.99
and 1.01. Variation of b, i.e. the nucleon size, does not affect
the results. But the size of EMC ratios is important   as x
becomes closer to 1. Our calculated neutron to proton structure
functions ratios are also in good agreement with present
theoretical calculation in which different models and
approximations have been used [17-22].

In conclusion, we have calculated the neutron to proton structure
functions ratios in the frame-work of quark-exchange model. We
have treated u and d quarks as well as proton and neutron
explicitly in our formalism and we have found satisfactory
results compare to present available data and others theoretical
models. However, we can improve our calculation by ({\it i})
treating the Fermi motion effect explicitly in the quark-exchange
frame-work i.e. calculating the expansion we made in equations
(13)-(16) as well as using other choice of nuclear wave-functions
which have been calculated with the new nucleon-nucleon
potentials, ({\it ii}) evaluating the connected three-body
diagram which we have been ignored in the present calculation
[15] and finally ({\it iii}) taking into the account the
nucleon-nucleon correlations. The later is interesting, because
for nuclei one could have non-vanishing structure functions for x
larger than one [29,30], which can be measured and calculated by
considering  the effect of short-range correlations in nuclei.

MM would like to thank  the University of Tehran  for supporting
him under the grants provided by its Research Council.


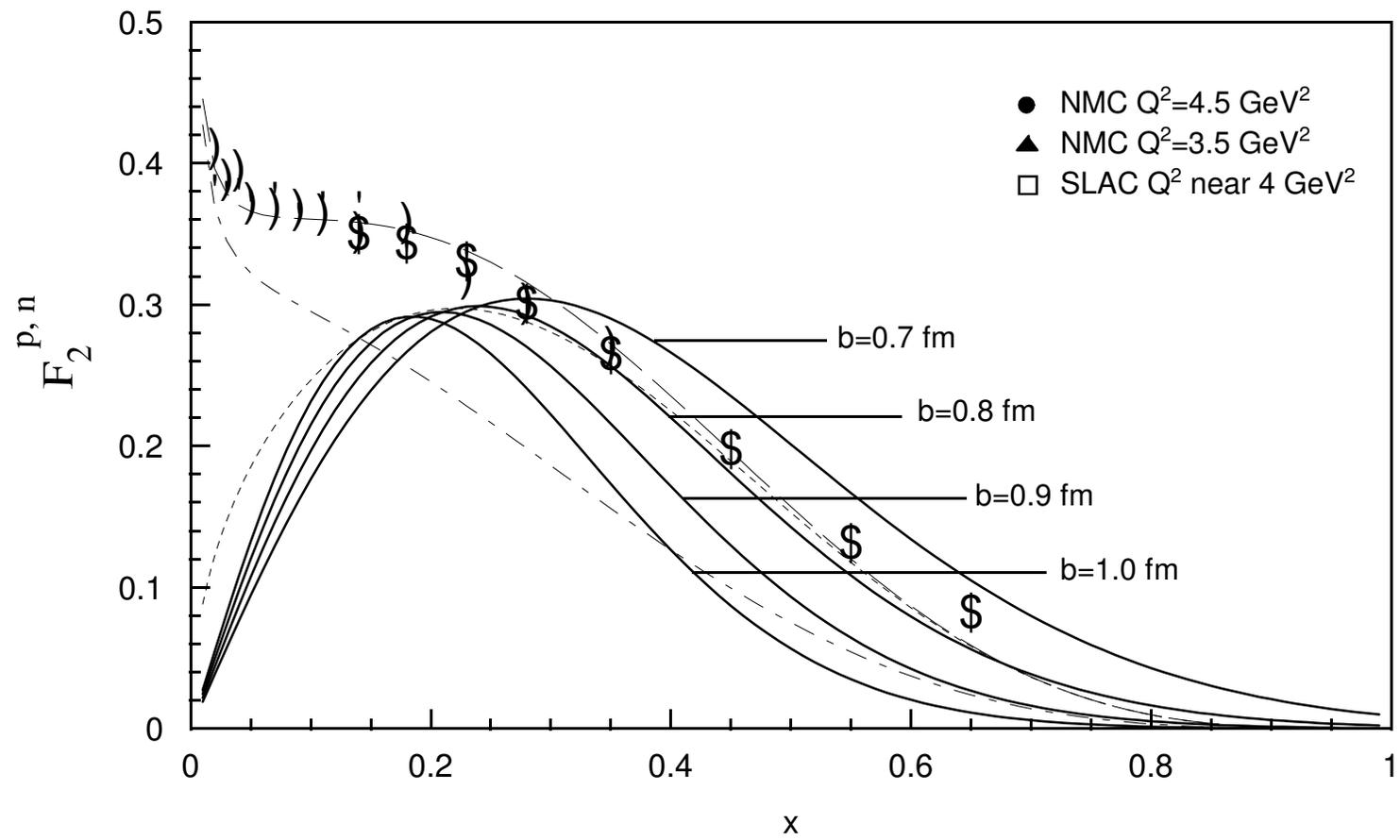
\begin{figure}
\caption{The proton structure function, only with valence quarks,
for $(m,\epsilon_0)$ pairs of $(165MeV,150MeV)$ and
$(165MeV,250MeV)$ at $Q^2=4 GeV^2$ and various b values. The pairs
$(165MeV,150MeV)$ and $(165MeV,250MeV)$ at $Q^2=4 GeV^2$ have been
chosen such that $b=0.8 fm$ gives the best fit to the GRV's
valence-quarks distributions (dotted curve)[25]. The dash and
dotted dash curves are the GRV's full NLO proton and neutron
distribution functions. The  NMC and SLAC data [19,28] for the
structure function of proton have been also given for comparison.}
\label{fig1}
\end{figure}
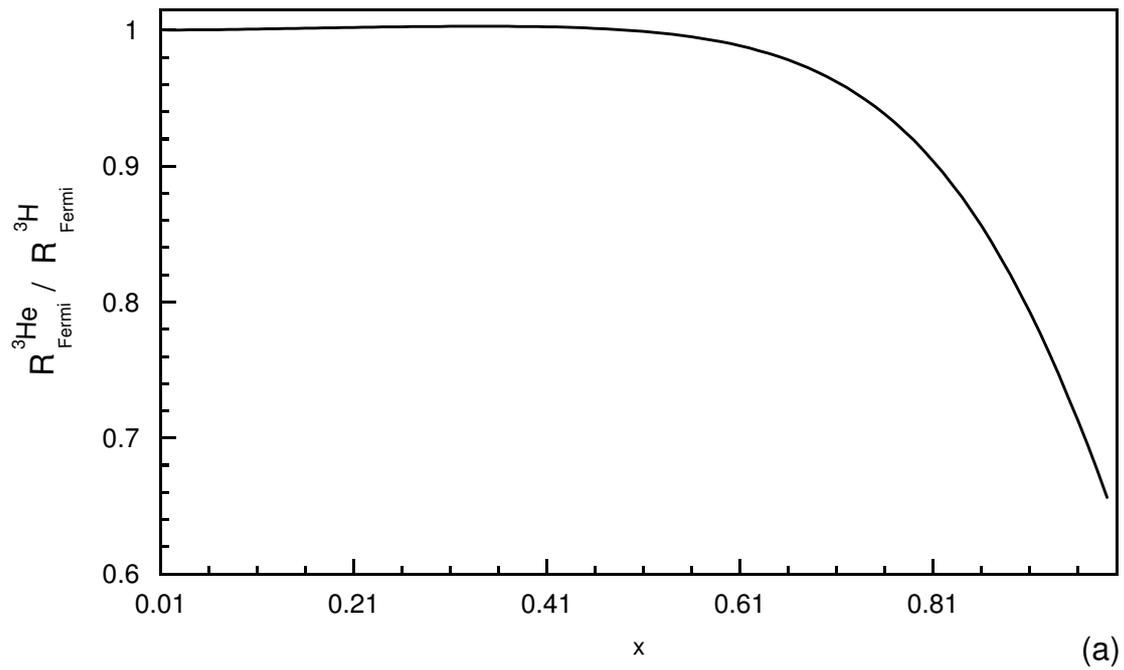
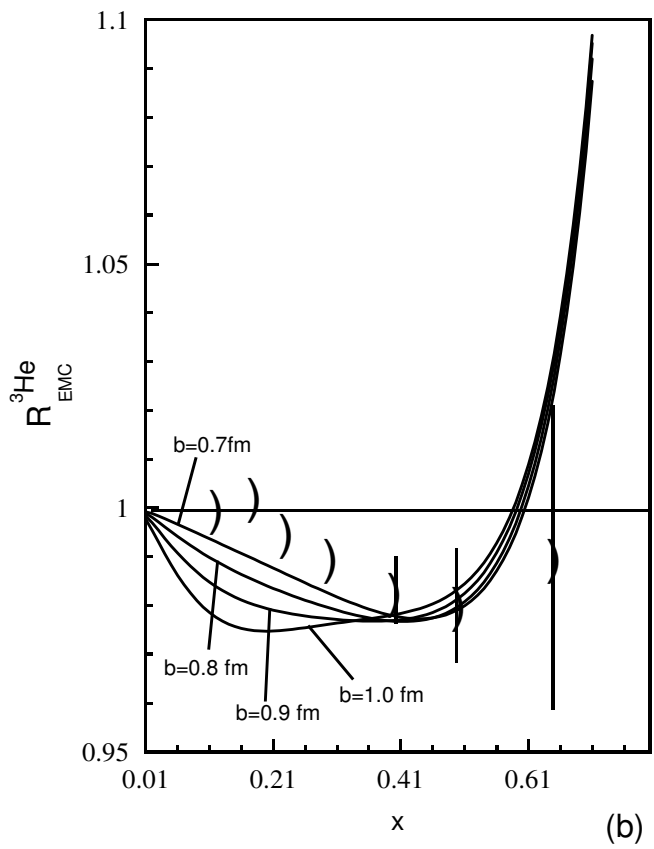
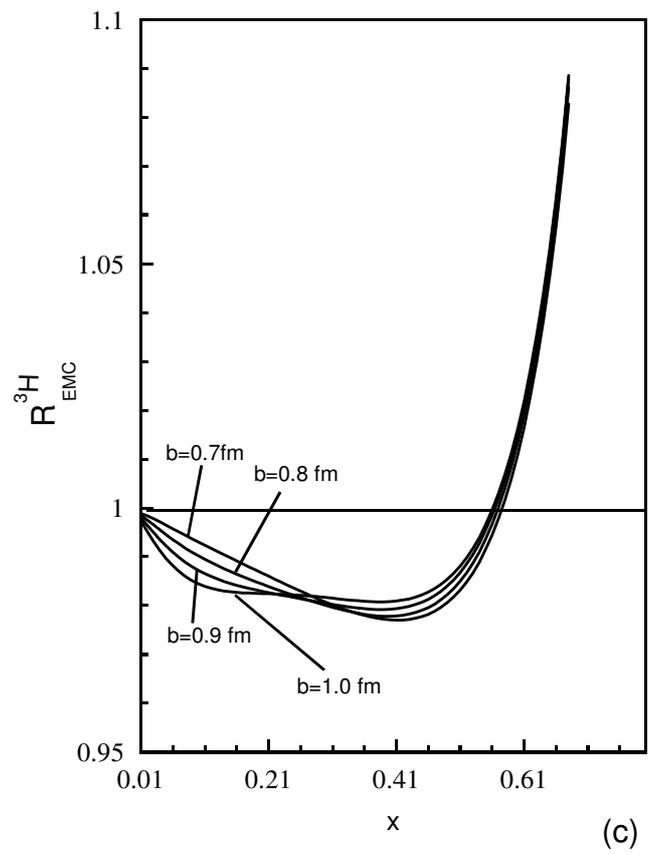
\begin{figure}
\caption{(a) The EMC ratios of the structure functions of helium 3
to tritium by considering only the Fermi motion effect with
parameters given in the text. The EMC effect in helium 3 (b) and
tritium (c) for different values of b.} \label{fig2}
\end{figure}
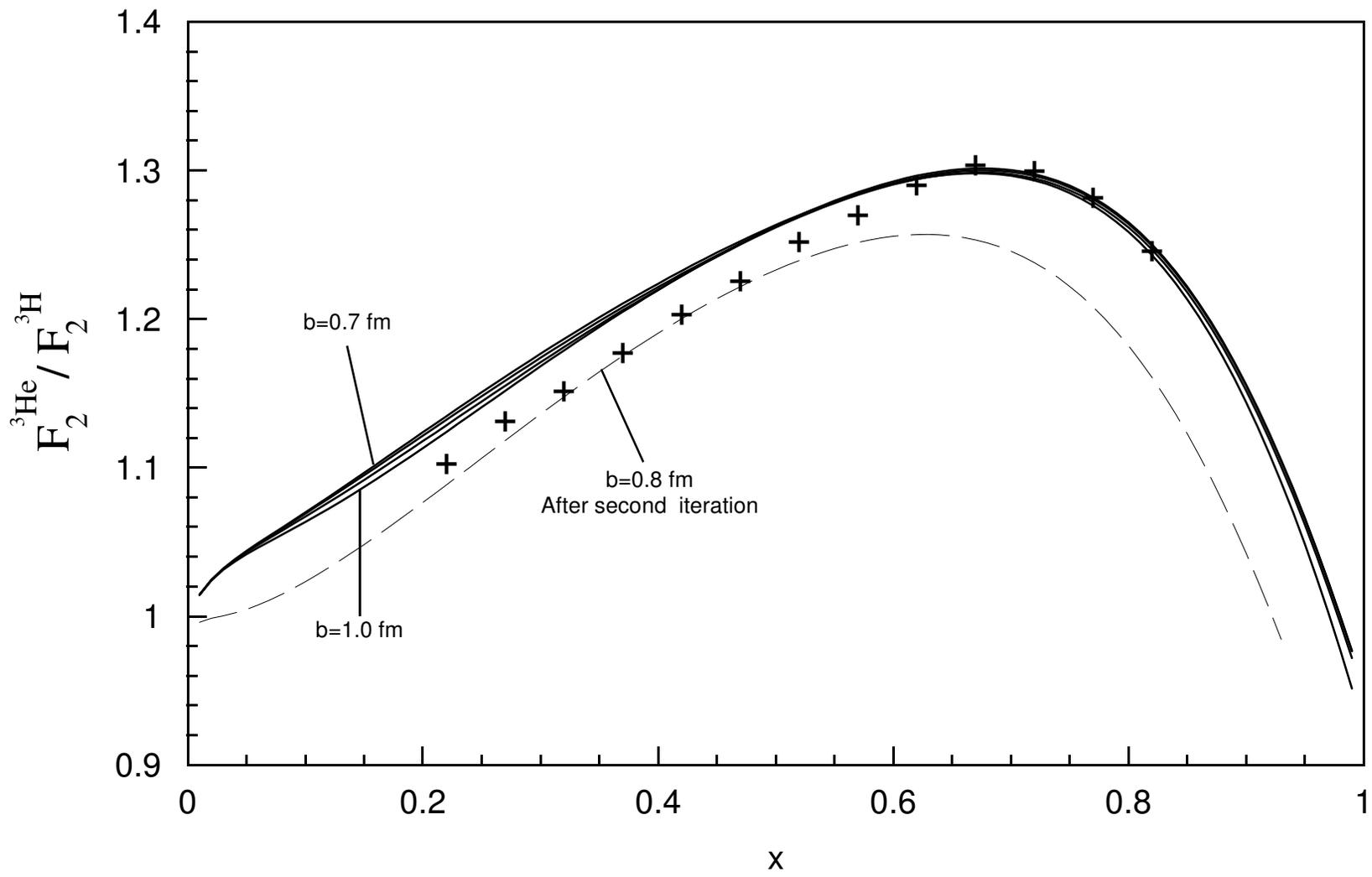
\begin{figure}
\caption{(a) The ratios of structure functions of helium 3 to
tritium (equation (29))for different values of b (full curves).
The dash curve is after second iteration with ${\cal
R}_{EMC}^{^3He/{^3H}}(x,Q^2)=1.01$ . The sun-burst point are from
references [10,20] (see also the text). } \label{fig3}
\end{figure}
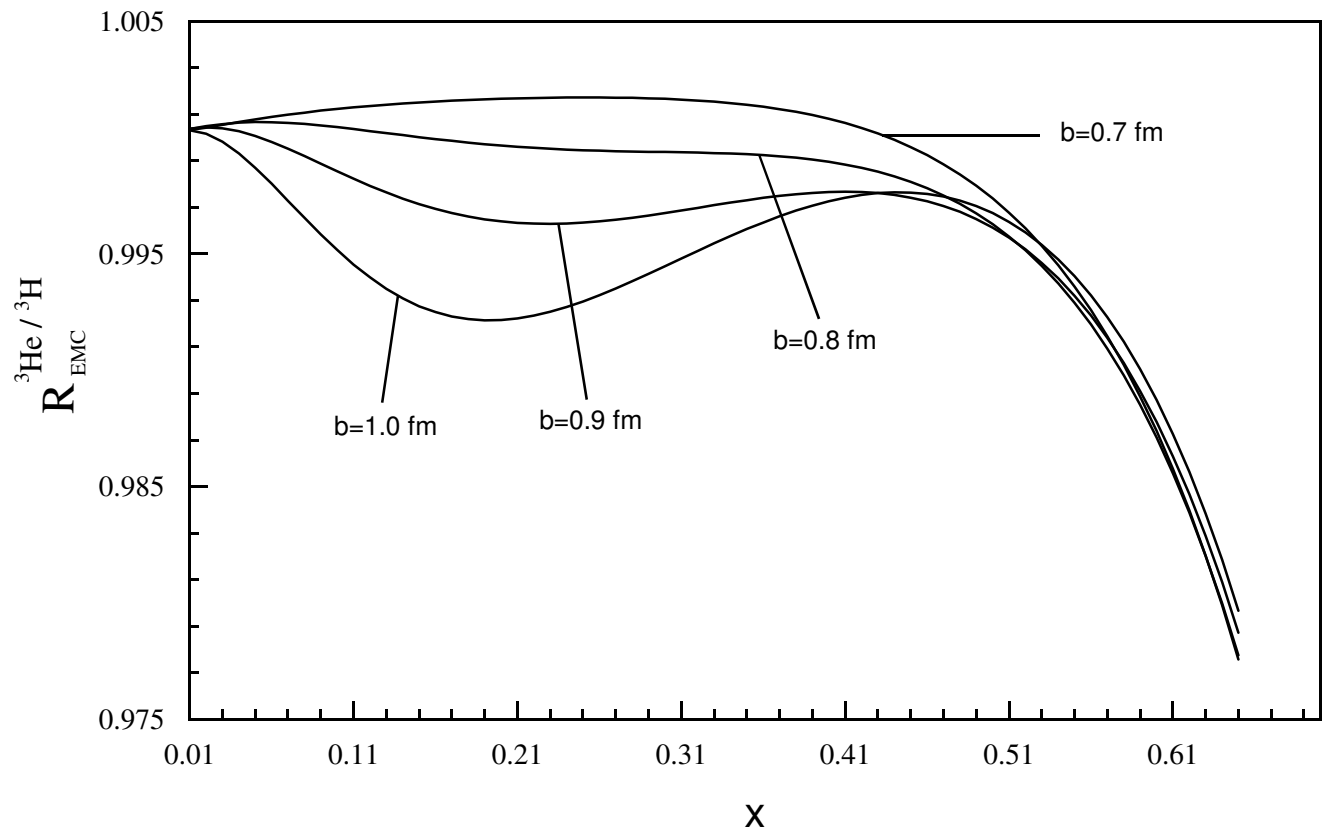
\begin{figure}
\caption{The  EMC ratios for different values of b without any
constrain (equation (28)).} \label{fig4}
\end{figure}
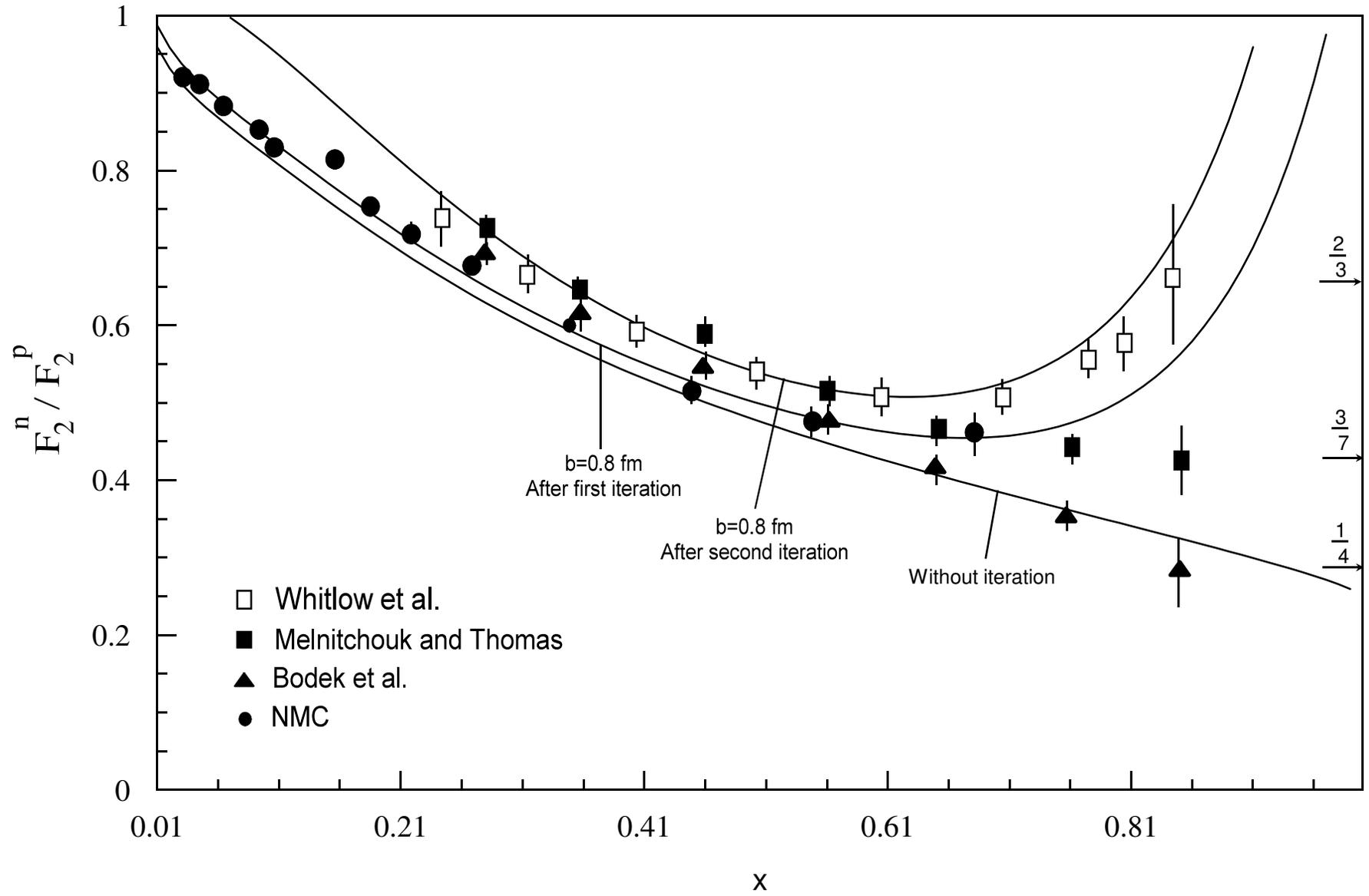
\begin{figure}
\caption{The neutron to proton structure functions ratios
(equation (31), see the text for more details ). The data are from
Whitlow et al. [19], Melnitchouk and Thomas [18], Bodek et al.
[17] and NMC [28].} \label{fig5}
\end{figure}
\end{document}